\documentclass[%
 reprint,
msmath,amssymb,
aps,
prx,
]{revtex4-1}
\pdfoutput=1
\usepackage{graphicx}
\usepackage{amsmath}
\usepackage{amssymb}
\usepackage{dsfont}
\usepackage{tensor}
\usepackage{color}
\usepackage{upgreek}
\usepackage{braket}
\usepackage{hyperref}
\usepackage[letterpaper,left=1.5cm,right=1.5cm,top=1.6 cm,bottom=2.5cm]{geometry}
\usepackage{float}
\usepackage[dvipsnames]{xcolor}
\definecolor{mygray}{gray}{1}
\usepackage[utf8]{inputenc}
\usepackage[normalem]{ulem}

\newcommand{\Ariel}{\textcolor{black}}

\usepackage[normalem]{ulem}

\begin{document}
\pdfoutput=1
\preprint{APS/123-QED}

\title{From the open Heisenberg model to the Landau-Lifshitz equation}

\author{Ariel Norambuena}
\affiliation{Centro de Investigaci\'on DAiTA Lab, Facultad de Estudios Interdisciplinarios, Universidad Mayor, Chile}
\author{Andr\'es Franco}
\affiliation{Centro de Investigaci\'on DAiTA Lab, Facultad de Estudios Interdisciplinarios, Universidad Mayor, Chile}
\author{Ra\'ul Coto}
\affiliation{Centro de Investigaci\'on DAiTA Lab, Facultad de Estudios Interdisciplinarios, Universidad Mayor, Chile}

\date{\today}

\begin{abstract}

Magnetic systems can be described by the classical Landau-Lifshitz (LL) equation or the fully quantum open Heisenberg model. Using the Lindblad master equation and the mean-field approximation, we demonstrate that the open Heisenberg model is reduced to a generalized LL equation. The open dynamic is modeled using spin-boson interactions with a common bosonic reservoir at thermal equilibrium. By tracing out the bosonic degrees of freedom, we obtain two different decoherence mechanisms: on-site dissipation and an effective spin-spin interaction mediated by bosons. Using our approach, we perform hysteresis calculations, closely connected with the Stoner-Wohlfarth theory. We compare the exact numerical master equation and the mean-field model, revealing the role of correlations originated by non-local interactions. Our work opens new horizons for the study of the LL dynamics from an open quantum formalism.

\end{abstract}

\pacs{Valid PACS appear here}
\maketitle

\section{Introduction}

Since its discovery in 1928~\cite{Heisenberg1928}, the original Heisenberg exchange interaction between two spins $J \mathbf{S}_1 \cdot \mathbf{S}_2$ has been extended to complex magnetic arrangements and successfully implemented in a variety of quantum systems. Nowadays, myriads of physical models describing the interaction between $N$ spin-$1/2$ particles are based on particular cases of the Heisenberg Hamiltonian

\begin{equation} \label{Hamiltonian}
H = \gamma \mathbf{B} \cdot \sum_{j=1}^{N}\mathbf{S}^{(j)}+\sum_{\alpha,\beta}\sum_{i \neq j}V_{\alpha \beta}^{ij} S_{\alpha}^{(i)}S_{\beta}^{(j)},
\end{equation}

where $\alpha, \beta = x,y,z$ are the components of the spin operators. A suitable choice of the magnetic field $\mathbf{B}$, the exchange coupling constants $V_{\alpha \beta}^{ij}$, and the topology of the system allows to understand the origin of magnetic ordering~\cite{Swendsen1974}, phase transition~\cite{Kawakami2001}, spin-wave excitations~\cite{Maksimov2019}, lattice effects~\cite{Andreas2009}, to name a few. Furthermore, the Heisenberg dynamic is currently reproduced in different physical systems such as circuit quantum electrodynamics~\cite{Wallraff2015}, cavity QED~\cite{Mivehvar2019}, superconducting devices~\cite{Deshui2019}, one-dimensional interacting spins~\cite{Dominic2016}, Rydberg atoms~\cite{Muller2008, Lee2012, Jurcevic2017,Nguyen2018}, and trapped ions~\cite{Lanyon2011,Friedenauer2008}. Many of the previous setups deal with unavoidable relaxation processes induced by system-bath interactions, which is well understood in terms of the Lindblad master equation~\cite{Breuerbook,RivasHuelgaBook}. We use the term \textit{open Heisenberg model} (OHM) to describe any system of $N$ interacting spins with a Hamiltonian structure similar to Eq.~\eqref{Hamiltonian} and subject to an interaction with an external bath~\cite{Deshui2019, Dominic2016, Ashrafi2014}. \par 

From the classical point of view, the dynamics of spins can be described by the Landau-Lifshitz (LL) equation~\cite{LL1935}

\begin{equation} \label{LLG}
{1 \over \gamma}{d \mathbf{M} \over dt} = -\mathbf{M} \times \mathbf{B}_{\rm eff} - {\alpha \over |\mathbf{M}|} \mathbf{M} \times \left( \mathbf{M} \times \mathbf{B}_{\rm eff}\right),
\end{equation}

where $\mathbf{M}$ is the magnetization of the system and $\mathbf{B}_{\rm eff}$ is an effective magnetic field which includes internal and external contributions to the magnetization dynamics of the system, such as the anisotropy and the applied field. When $\alpha = 0$ the magnetization undergoes an endless precessional motion around the axis $\hat{n} = \mathbf{B}_{\rm eff}/|\mathbf{B}_{\rm eff} |$. However, the term $-\alpha\mathbf{M} \times \left( \mathbf{M} \times \mathbf{B}_{\rm eff} \right)$ introduces a phenomenological damped movement that preserves the magnitude of $\mathbf{M}$ leading to a stationary state fixed in time and parallel to the axis $\hat{n}$. Historically, Eq.~\eqref{LLG} was initially proposed by Landau and Lifshitz in 1935~\cite{LL1935} and later modified by Gilbert as the Landau-Lifshitz-Gilbert (LLG) equation~\cite{Gilbert} $\dot{\mathbf{M}}/\tilde{\gamma} =- \mathbf{M}\times \mathbf{B}_{\rm eff} + \tilde{\alpha} \mathbf{M}\times \dot{\mathbf{M}}/|\mathbf{M}|$ to better account for the effects of strong damping $\alpha$. Both formalisms (LL and LLG) are equivalent when $\gamma=\tilde{\gamma} / \left(1 + \tilde{\gamma}^2 \tilde{\alpha}^2 |\mathbf{M}|^2\right)$ and $\alpha/|\mathbf{M}| =\tilde{\gamma} \tilde{\alpha} / \left(1 + \tilde{\gamma}^2 \tilde{\alpha}^2 |\mathbf{M}|^2\right)$~\cite{AharoniBook}, and are extensively used for theoretical calculations of the dynamics of magnetic systems~\cite{Lakshmanan2011,Kambersky1975,Kalinikos1986,Rama2020,Lars2017,Michael2015}. Over the last decades these equations (and its variations) have proven to be an indispensable and versatile tool to describe a wide range of phenomena, like ferromagnetic resonance~\cite{Kittel1948}, propagation of spin waves~\cite{Burkard1990,Kalinikos1986,Camley1997}, spin transfer and spin-orbit torques~\cite{Slonczewski1996,Berger1996,Stiles2002,Tingsu2016,Manchon2009,Garate2009,Ioan2011}, temperature dynamics~\cite{Garanin1997,Palacios1998,Atxitia2017}, nuclear magnetic resonance~\cite{Bloch1946, Bloembergen1950} or conducting ferromagnets~\cite{Zhang2009}, among others. \par

Because of the universality of the Heisenberg model to describe magnetic properties, the following question arises. Can the OHM reproduce magnetization dynamics similar to the classical LL equation? The answer is yes, and more importantly, we shall demonstrate that the dynamics is more general under certain conditions. Here, we establish an unexplored connection between the LL equation and a particular Lindblad superoperator in a 
Markovian master equation. It is worth noticing that other quantum approaches have successfully addressed the microscopic derivation of the LL equation by considering quantum processes. For instance, using the spin-wave theory~\cite{Herbert1958}, the Fokker-Planck equation~\cite{Garanin1997}, the Dirac-Kohn-Sham theory~\cite{Mondal2016,Mondal2018}, a mean-field tight-binding model~\cite{Felipe2019}, a non-Hermitian Hamiltonian approach~\cite{Wieser2015}, and the Yang-Mills equation~\cite{Naoto2019}. Nevertheless, our approach is a new perspective that is useful for connecting the LL dynamics with the evolution of open quantum systems in the mean-field regime. The latter is particularly advantageous to future simulations of magnetic-like phenomena using quantum systems like trapped ions~\cite{Lanyon2011,Friedenauer2008}, superconducting devices~\cite{Deshui2019} and cavity QED~\cite{Mivehvar2019}. Moreover, 
it allows the study of more general environments exhibiting memory effects usually described as non-Markovian master equations~\cite{Vega2017}. \par

\begin{figure*}[ht]
\centering
\includegraphics[width=0.8 \linewidth]{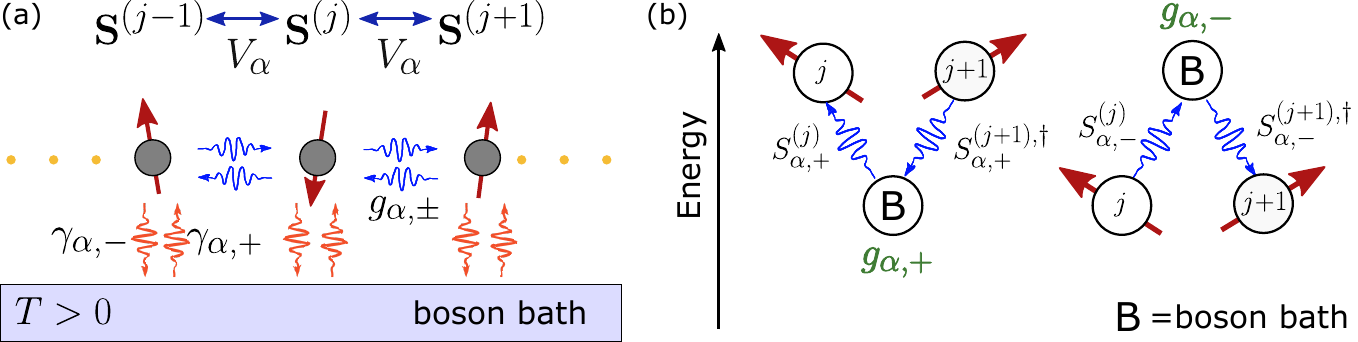}
\caption{(a) Schematic representation of the effective open Heisenberg model. Each spin $\mathbf{S}^{(j)}$ is coupled to its nearest-neighbour via the coupling constants $V_{\alpha}$. The common boson bath allows the on-site decay channels $\gamma_{\alpha, \pm}$ and the dissipative hopping terms $g_{\alpha, \pm}$. (b) Diagram showing the physical interpretation of the Lindbladian $\mathcal{L}_{\rm nn}(\rho)$~\eqref{MasterEquation}. Adjacent spins in the sites $(j+1)$ and $(j)$ exchange energy with the boson bath and tracing out over the bath degrees of freedom we obtain the effective dissipation $g_{\alpha,\pm}$.}
\label{fig:Figure1}
\end{figure*} 

The paper is organized as follows. In Sec.~\ref{Model}, we introduce the Hamiltonian of the system and the Markovian master equation. Section~\ref{MeanField_LL} introduces the mean-field approximation leading to a generalized version of the LL equation for the magnetization dynamics. Here, we discuss the hysteresis of the system in analogy with the Stoner-Wohlfarth theory~\cite{Stoner1948}. Finally, in Sec.~\ref{Density-Matrix} we numerically solve the quantum master equation to compare our results with the mean-field model. Finally, we discuss the effect of correlations by considering both closed and open dynamics for the isotropic and anisotropic Heisenberg models. \par

\section{Model} \label{Model}

We consider a linear spin chain composed of $N$ spin-$1/2$ particles with on-site and hopping interaction terms described by the following system Hamiltonian

\begin{equation} \label{SystemHamiltonian}
H_s = \gamma \mathbf{B} \cdot \sum_{j=1}^{N}\mathbf{S}^{(j)}+\sum_{\alpha}\sum_{j=1}^{N-1}V_{\alpha} S_{\alpha}^{(j)}S_{\alpha}^{(j+1)},
\end{equation}

where $\gamma$ is the electronic gyromagnetic ratio and $V_{\alpha}$ is the coupling between adjacent spins along the cartesian directions $\mathbf{e}_\alpha$. Here, $\mathbf{S}^{(j)} = (\hbar/2)\boldsymbol{\sigma}^{(j)}$ is the spin operator of the $j$-th particle, where $\sigma^{(j)}_{\alpha}$ are the Pauli matrices for $S=1/2$. Note that for $\mathbf{B} = B_x \mathbf{e}_{x}$ and $V_y = V_z = 0$, Eq.~\eqref{SystemHamiltonian} reduces to the standard transverse-field Ising model~\cite{Ates2012, Jurcevic2017, Dominic2016}. \par

Other relevant magnetic coupling terms such as the dipole-dipole (DD) or the Dzyaloshinskii-Moriya (DM) $\sum_{ij} D_{ij} \mathbf{S}^{(i)} \times \mathbf{S}^{(j)}$~\cite{Dzyaloshinsky1958,Moriya1960} are crucial to model magnetic defects in solid-state systems~\cite{Raul2020} or skyrmions~\cite{Rohart2013,Jiadong2011}, respectively. However, in this work, we focus either on small systems of individual spins, or large systems of homogeneous magnetic moments. In the former, neglecting DD interaction is valid as it is several orders of magnitude smaller than the exchange coupling, and becomes relevant only as a long-range interaction~\cite{Kittel1449,Janickaa2007}. Moreover, for large homogeneous systems, the DD interaction can be considered as an additional anisotropy~\cite{Andrew1993, Amikam1998}, and thus not calculated explicitly. Similarly, the DM interaction is typically present in magnetic systems with broken inversion symmetry~\cite{Giovanni2016,Jaehun2015,Moon2013}, which lies outside of the scope of this paper.  \par

In general, quantum systems interact with its surrounding environment, which originates from different damping mechanisms. In our model, we assume that the spin chain is coupled to a generic boson reservoir that is in thermal equilibrium. In order to introduce dissipative effects, we focus on Markovian evolutions~\cite{Breuerbook}, i.e., without memory effects. Non-Markovian dynamics have been analytically solved for some particular spin systems~\cite{Jing2018,Sagnik2019}. However, we are interested in the dynamical properties of the system at time scales larger than the characteristic bath correlation time, i.e., where the first and second Markov approximations hold~\cite{Vega2017}. Hence, dissipation originates from the following system-bath interaction Hamiltonian

\begin{equation}\label{Interaction}
V = \sum_{k}\sum_{j=1}^{N}\sum_{\alpha}\left[g_{\alpha jk} S_{\alpha,+}^{(j)}a_k +  g_{\alpha jk}^{\ast}S_{\alpha,-}^{(j)}a_k^{\dagger}\right],
\end{equation}

where $k$ runs over infinite bosonic modes of the environment and $g_{\alpha jk}$ are the spin-bath coupling constants. The environment is considered as a collection of harmonic oscillators described by the bath Hamiltonian $H_b = \sum_k \omega_k a_k^{\dagger}a_k$, where $a_k$ and $a_k^{\dagger}$ are the annihilation and creation boson operators, respectively. Spin operators $S_{\alpha,\pm}^{(j)}$ are defined as

\begin{eqnarray}
S_{x,\pm}^{(j)} &=& -S_z^{(j)} \pm \mbox{i} S_y^{(j)},\\
S_{y,\pm}^{(j)} &=& S_z^{(j)} \pm \mbox{i} S_x^{(j)}, \\
S_{z,\pm}^{(j)} &=& S_x^{(j)} \pm \mbox{i} S_y^{(j)},
\end{eqnarray}

with $S_{\alpha,+}^{(j)} \ket{\downarrow}_{\alpha}^{(j)} = \ket{\uparrow}_{\alpha}^{(j)}$ (raising) and $S_{\alpha,-}^{(j)} \ket{\uparrow}_{\alpha}^{(j)} = \ket{\downarrow}_{\alpha}^{(j)}$ (lowering) describing spin flip-flop processes between the eigenstates ($ \ket{\downarrow}_{\alpha}^{(j)}, \ket{\uparrow}_{\alpha}^{(j)}$) of the Pauli matrices $\sigma_{\alpha}^{(j)}$. The spin states $\ket{\uparrow}_{\alpha}^{(j)}$ (excited) and $\ket{\downarrow}_{\alpha}^{(j)}$ (ground) are given in Eqs.~\eqref{xup}-\eqref{zdown}. We remark that the interaction Hamiltonian~\eqref{Interaction} assumes that all spins couple to the same environment. As a consequence, spins will experience an effective coupling to each other in the Markovian master equation~\cite{Michael2001} [In our calculations it will appear as the last term in Eq.~\eqref{MasterEquation}]. The interaction Hamiltonian~\eqref{Interaction} can be written as

\begin{eqnarray}\label{NewFormInteraction}
V &=& \sum_{k}\sum_{j=1}^{N}\left[S_z^{(j)}d_k\left(a_k +  a_k^{\dagger}\right) + 
\left(e_k S_+^{(j)}a_k +  e_k^{\ast} S_-^{(j)}a_k^{\dagger}\right) \right] +\nonumber\\
&&+\sum_{k}\sum_{j=1}^{N}\left(c_k S_-^{(j)}a_k +  c_k^{\ast} S_+^{(j)}a_k^{\dagger}\right),
\end{eqnarray}

where $d_k = g_{yjk}-g_{xjk}$, $e_k = g_{xjk}+ig_{yjk}+g_{zjk}$, $c_k = -g_{xjk}+ig_{yjk}$, and $S_{\pm}^{(j)} = S_x^{(j)} \pm i S_y^{(j)}$. \Ariel{Eq.~\eqref{NewFormInteraction}} reveals that the interaction Hamiltonian~\eqref{Interaction} \Ariel{is composed by} pure-dephasing ($d_k$), \Ariel{amplitude damping} or \textit{energy-exchange} ($e_k$) and counter-rotating terms ($c_k$). For $g_{xjk} = g_{yjk} = 0$ the interaction Hamiltonian reduces to the standard amplitude damping model~\cite{Santos2014}, where bosons are absorbed or emitted inducing spin flip-flop processes between the states $\ket{\downarrow}_{z}^{(j)}$ and $\ket{\uparrow}_{z}^{(j)}$. The full Hamiltonian $H = H_s + H_b + V$ \Ariel{is similar to the} \textit{chain-boson model}~\cite{Skinner2008}, however our interaction is more general since we are including pure-depashing and counter-rotating terms.\par  

In the Markov and secular approximations, we derive the following master equation that considers effective spin interactions up to first nearest-neighbors

\begin{eqnarray}
{d \rho \over dt} &=& -i[H_s,\rho] \nonumber \\
&& +\sum_{j=1}^{N}\sum_{\alpha}\sum_{\eta = \pm} \gamma_{\alpha,\eta}\left[S_{\alpha,\eta}^{(j)} \rho S_{\alpha,\eta}^{(j),\dagger} -{1 \over 2}\left\{S_{\alpha,\eta}^{(j),\dagger}S_{\alpha,\eta}^{(j)}, \rho \right \} \right] \nonumber \\
&& +\sum_{\langle j,j'\rangle}^{}\sum_{\alpha}\sum_{\eta = \pm} g_{\alpha,\eta}\left[S_{\alpha,\eta}^{(j')} \rho S_{\alpha,\eta}^{(j),\dagger} -{1 \over 2}\left\{S_{\alpha,\eta}^{(j),\dagger}S_{\alpha,\eta}^{(j')}, \rho \right \} \right], \nonumber \\  
 &=& -i[H_s,\rho] + \mathcal{L}_{\rm os}(\rho) + \mathcal{L}_{\rm nn}(\rho), \label{MasterEquation}
\end{eqnarray}

where $\gamma_{\alpha,+}$ and $\gamma_{\alpha,-}$ are damping rates associated to absorption and emission processes between each site and the external boson environment, respectively. Here, $\langle j,j'\rangle$ denotes nearest-neighbor spins by considering all terms satisfying the condition $|j-j'|=1$. The \textit{on-site Lindbladian} $\mathcal{L}_{\rm os}(\rho)$ has already been implemented using Rydberg atoms at low temperatures~\cite{Lee2012,Muller2008}, with $\gamma_{z,-} > 0$ and $\gamma_{x,\eta} = \gamma_{y,\eta} = \gamma_{z,+} = 0$. On the other hand, we called the superoperator $\mathcal{L}_{\rm nn}(\rho)$ as the \textit{nearest-neighbours Lindbladian} since it accounts for the effective energy-exchange between adjacent spins. This dissipation's source naturally appears in the master equation of multi-atomic systems coupled to light~\cite{Ficek2005}, and it is a pivotal result towards connecting the OHM  with the LL theory, since $\mathcal{L}_{\rm nn}(\rho)$ will reproduce the damping term $\mathbf{M}\times(\mathbf{M} \times \mathbf{B}_{\rm eff})$ in equation~\eqref{LLG}. Further details of the microscopic derivation of the master equation~\eqref{MasterEquation} is presented in Appendix~\ref{appendix2}, and it follows the spirit of the open dynamics for interacting qubits presented in Ref.~\cite{Santos2014}. A representation of the OHM is depicted in Fig.~\ref{fig:Figure1}. \par

In the next section, we use the mean-field approximation to solve the OHM given in Eq.~\eqref{MasterEquation}, and after introducing the non-linear single-particle dynamics, we show its connection with the LL Eq.~\eqref{LLG}.

\section{mean-field approximation and Landau-Lifshitz equation}\label{MeanField_LL}

The mean-field approximation considers the many-body density matrix of the $N$-particle system as a separable tensor product of single-particle density matrices $\rho_j(t)$, such that

\begin{equation}\label{MeanFiedlState}
\rho_{\rm MF}(t) = \rho_{1}(t) \otimes \cdots \otimes \rho_N(t).
\end{equation}

The above factorization is more accurate as $N$ increases~\cite{Merkli2012} and it has been also discussed in the context of open quantum systems~\cite{Breuerbook,Spohn1980} or using the name of Hartree approximation~\cite{Bonitz1998}. Each density operator in Eq.~\eqref{MeanFiedlState} is described as a two-level system, where $\rho_j(t) = (1/2)(\mathds{1}+\mathbf{f}(t) \cdot \boldsymbol{\sigma}^{(j)})$, with $\mathbf{f}(t) = f_x(t)\mathbf{e}_x+f_y(t)\mathbf{e}_y+f_z(t)\mathbf{e}_z$ and $\boldsymbol{\sigma}^{(j)} = \sigma_x^{(j)} \mathbf{e}_x+\sigma_y^{(j)}\mathbf{e}_y+\sigma_z^{(j)} \mathbf{e}_z$. To shed more light on the \Ariel{magnetization} dynamics of the system we introduce the magnetic moment of each particle through the relation $\boldsymbol{\mu}^{(j)} = -g_s \mu_B \mathbf{S}^{(j)}/\hbar$, where $\mu_B = 9.27 \times 10^{-24}$ JT$^{-1}$ is the Bohr magneton, and $g_s \approx 2$ is the $g$-factor. After averaging the effect of all magnetic moments the macroscopic magnetization reads

\begin{equation} \label{Magnetization}
\mathbf{M} = {1 \over N}\sum_{j=1}^{N}{1 \over V} \langle \boldsymbol{\mu}^{(j)} \rangle = {\mu_B \over V N} \sum_{j=1}^{N} \langle \boldsymbol{\sigma}^{(j)} \rangle,
\end{equation}

where $N$ is the number of spins, $V = L^3$ is a characteristic volume of the system, $L$ is a characteristic length, and $\langle \sigma^{(j)}_{\alpha}\rangle = \mbox{Tr}[\sigma^{(j)}_{\alpha}\rho_{j}]$ is the expectation value of each Pauli operator. In what follows we focus on the particular case $\gamma_{x,\pm} = \gamma_{y,\pm} = 0$ in order to reduce the number of damping rates in our analysis. Thus, we can explicitly calculate the single-particle dynamics by solving $d\rho_2/dt = \mbox{Tr}_{1,3,...,N}[\mathcal{L}(\rho_{\rm MF})]$, where $\mbox{Tr}_{1,3,...,N}$ is the partial trace over the remaining $N-1$ particles (without considering $j=2$). After applying the partial trace, we obtain the following set of coupled non-linear equations:


\begin{widetext}
\begin{eqnarray}
{dM_x \over dt}&=& M_z\left(\gamma B_y + m_y V_{y}\right)-M_y\left( \gamma B_z + m_z V_z\right) + {1 \over 2 |\mathbf{M}|}\left[-g_{z} M_x M_z - g_{y} M_x M_y + g_{x}(M_y^2+M_z^2)\right]  - {1 \over 2}\Gamma M_x, \label{x}\\
{dM_y \over dt} &=& M_x\left(\gamma B_z + m_z V_z\right)-M_z\left(\gamma B_x + m_x V_{x}\right) + {1 \over 2 |\mathbf{M}|}\left[-g_{z} M_y M_z - g_{x} M_x M_y + g_{y}(M_x^2+M_z^2)  \right] - {1 \over 2}\Gamma M_y, \label{y}\\
{dM_z \over dt} &=& M_y\left(\gamma B_x + m_x V_{x}\right)-M_x\left(\gamma B_y + m_y V_{y}\right)+{1 \over 2 |\mathbf{M}|}\left[-g_{y} M_y M_z - g_{x} M_x M_z + g_{z}(M_x^2+M_y^2) \right]- \Gamma (M_z+1), \label{z} 
\end{eqnarray}
\end{widetext}

where $m_\alpha = M_\alpha/|\mathbf{M}|$, $\mathbf{M} = M_x \mathbf{e}_x+M_y \mathbf{e}_y+M_z \mathbf{e}_z$, $g_{\alpha} = g_{\alpha,-}-g_{\alpha,+}$, and $\Gamma = \gamma_{z,-} + \gamma_{z,+}$ is the total damping when $\gamma_{x,\pm} = \gamma_{y,\pm} = 0$. Because of the boundary conditions of the linear chain, the first ($j=1$) and last ($j=N$) particles interact with a single neighbor spin, in contrast to the intermediate spins ($j \neq 1,N$) that interact with two nearest-neighbors. Therefore, for particles $j=1,N$ Eqs.~\eqref{x}-\eqref{z} must be modified by considering $V_{\alpha} \rightarrow V_{\alpha}/2$. For $\mathbf{B} = B_x \mathbf{e}_{x}$, $g_{\alpha} = 0$, and $V_x = V_y = 0$ the set of non-linear Eqs.~\eqref{x}-\eqref{z} reduces to the transverse-field open Ising model presented in Ref.~\cite{Dominic2016}. As expected, the single-particle dynamics is affected by the presence of an effective field induced by the other $N-1$ particles. In fact, the first two terms on the right-hand side of Eqs.~\eqref{x}-\eqref{z} are recognized as the effective magnetic field whose components are $B_{\rm eff,\alpha} =  \gamma B_\alpha + m_\alpha V_\alpha$. The third term on the right-hand side of Eqs.~\eqref{x}-\eqref{z} is crucial for the LL theory and for illustration, we write it in a compact form,

\begin{eqnarray}
L_x &=& -g_{z} M_x M_z - g_{y} M_x M_y + g_{x}(M_y^2+M_z^2), \\
L_y &=& -g_{z} M_y M_z - g_{x} M_x M_y + g_{y}(M_x^2+M_z^2), \\
L_z &=& -g_{y} M_y M_z - g_{x} M_x M_z +g_{z}(M_x^2+M_y^2).
\end{eqnarray}

By a direct calculation we get the constraint $\sum_{\alpha}M_{\alpha}L_{\alpha} = M_x L_x + M_y L_y + M_z L_z = 0$ which plays an important role on the dynamics, since it leaves invariant the magnitude of the magnetization vector $\mathbf{M}$. To understand this, we write the magnetization dynamics induced by $L_{\alpha}$ as $\dot{M}_{\alpha} = L_{\alpha}/2|\mathbf{M}|$. We note that $\sum_{\alpha} \dot{M}_{\alpha}M_{\alpha} = (1/2|\mathbf{M}|)\sum_{\alpha}M_{\alpha}L_{\alpha} = 0$. Thus, $\sum_{\alpha} \dot{M}_{\alpha}M_{\alpha} = (1/2)(d /dt)\sum_{\alpha}M_{\alpha}^2 = 0$, i.e. $d|\mathbf{M}|^2/dt = 0$. Additionally, the last term in Eqs.~\eqref{x}-\eqref{z} accounts for on-site dissipation induced by the boson bath that directly affects the magnitude of the magnetization vector. Based on these observations, and considering a large number of spins, we obtain the following dynamical equation for the macroscopic magnetization vector

\begin{equation} \label{MagnetizationDynamics1}
{1 \over \gamma} {d \mathbf{M} \over dt} = - \mathbf{M} \times \mathbf{B}_{\rm eff} - {1 \over |\mathbf{M}|}\mathbf{M} \times \left(\mathbf{M}\times \mathbf{D}\right) - \mathds{R} \mathbf{M} - \mathbf{R}_0,
\end{equation}

where $\mathbf{B}_{\rm eff} = \mathbf{B} + \textbf{B}_{\rm an}$ is the effective magnetic field responsible for the gyromagnetic precession of the magnetization vector. In our model, $\textbf{B}_{\rm an} = m_x V_x \mathbf{e}_x+ m_yV_y \mathbf{e}_y + m_z V_z \mathbf{e}_z$ is the anisotropy field caused by the local interaction between spins~\cite{Garanin1997} [see Heisenberg Hamiltonian~\eqref{SystemHamiltonian}]. On the other hand, we recognize $\mathbf{D} = (1/2 \gamma)(g_{x}\mathbf{e}_x + g_{y} \mathbf{e}_y + g_{z} \mathbf{e}_z)$ as the magnetic field responsible for modifying the precession of the magnetization similar to LL equation~\eqref{LLG}. The last two terms in Eq.~\eqref{MagnetizationDynamics1} are given by

\begin{equation}
\mathds{R} =  \left( \begin{array}{ccc}
                     \Gamma/2 & 0 & 0 \\
                     0 &  \Gamma/2 & 0 \\
                     0 & 0 & \Gamma \\
                   \end{array} \right), \quad 
                   \mathbf{R}_0 = \left( \begin{array}{c}
                    0 \\
                    0 \\
                    \Gamma 
                   \end{array}\right), 
\end{equation}

where $\mathds{R}$ is the relaxation tensor~\cite{Junk} and $\mathbf{R}_0$ is the noise induced by the boson environment. These terms are in agreement with the Bloch theory applied to magnetic systems~\cite{Bloch1946}. 

\begin{figure*}[ht]
\centering
\includegraphics[width=1 \linewidth]{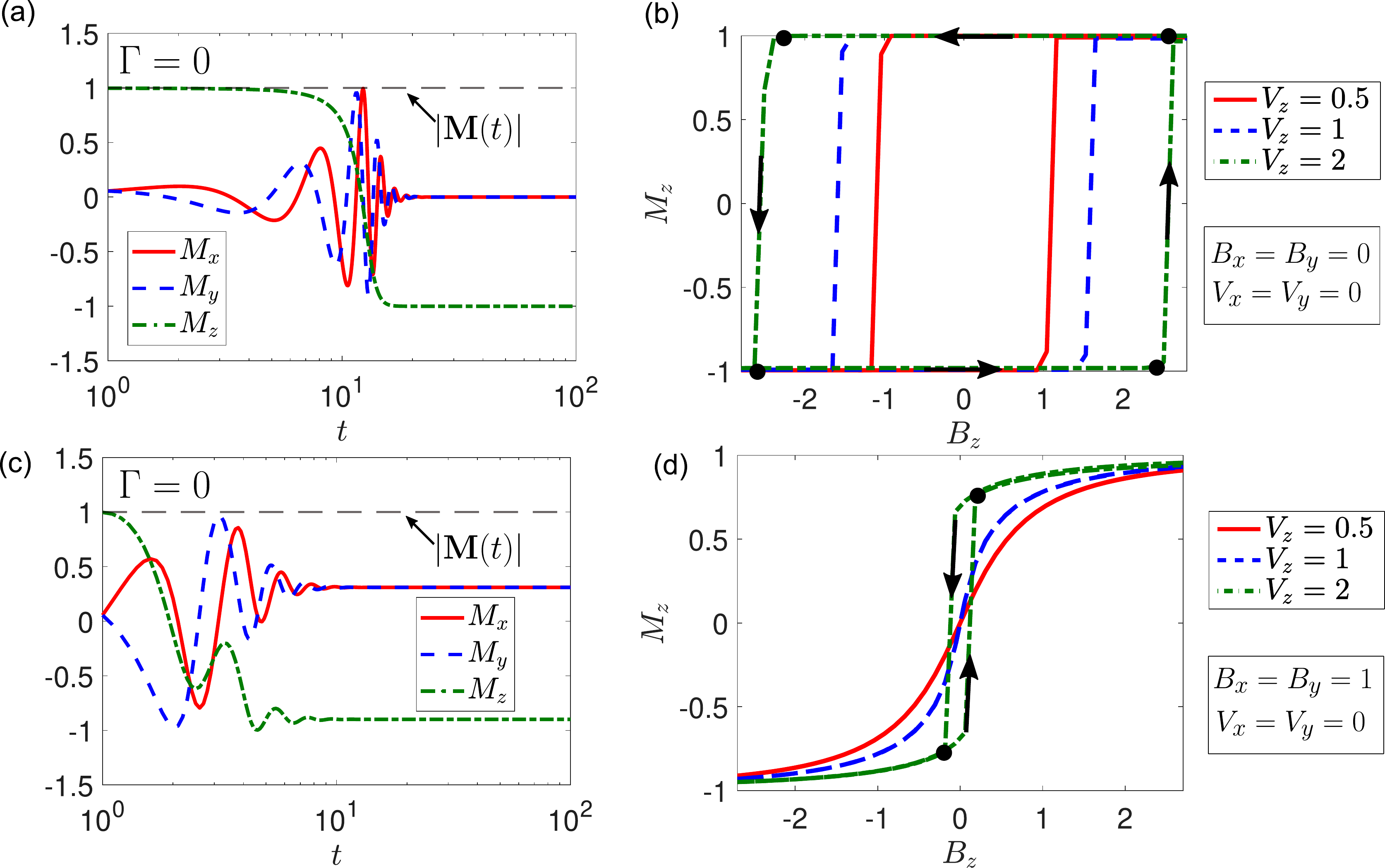}
\caption{(a) Numerical solution for the magnetization components $M_{\alpha}(t)$ in the mean-field approximation starting from the initial condition $\mathbf{M}(0) = |\mathbf{M}(0)|(\cos(\phi_0)\sin(\theta_0),\sin(\phi_0)\sin(\theta_0),\cos(\theta_0))$ with $\theta_0 = \pi/40$, $\phi_0 = \pi/4$, and $N = 500$. We fix $\Gamma = B_x = B_y = V_x = V_y = 0$, $\alpha = 0.5$, $V_z = 1$, and $B_z = -2$. The gray dashed line shows the magnitude $|\mathbf{M}(t)|$ which is constant when $\Gamma = 0$. (b) Hysteresis curve in the mean-field approximation for different values of the coupling $V_z$. (c) Numerical solution for the magnetization components using $B_{x,y} = 1$, $B_z = -2$, $V_{x,y} = 0$, $V_z = 0.5$ and $\alpha = 0.5$. (d) Effect of the $B_{x,y}$ components on the hysteresis curve for different values of the coupling $V_z$.}
\label{fig:Figure2}
\end{figure*}

In order to illustrate the scope of Eq.~\eqref{MagnetizationDynamics1}, we derive some well-known models as particular cases. First, for $\mathbf{D} = \alpha \mathbf{B}_{\rm eff}$ and $\Gamma = 0$, Eq.~\eqref{MagnetizationDynamics1} exactly reduces to the LL equation~\eqref{LLG}, where $\alpha$ is the dimensionless damping factor~\cite{LL1935,Gilbert}. Second, for $\mathbf{D} = \mathbf{0}$, our model is reduced to

\begin{equation}
{1 \over \gamma} {d \mathbf{M} \over dt} = - \left(\mathbf{M} -\mathbf{M}^{\rm SS} \right) \times \mathbf{B}_{\rm eff} - \mathds{R}\left(\mathbf{M} -\mathbf{M}^{\rm SS} \right),
\end{equation}

which is known as the \textit{Bloch–Bloembergen equation}
~\cite{Bloch1946, Bloembergen1950}, with $\mathbf{M}^{\rm SS}$ being the stationary state. Finally, for $\mathbf{R}_0 = 0$ Eq.~\eqref{MagnetizationDynamics1} reduces to the \textit{Callen's equation}~\cite{Rene2000,Callen1958}, which is a phenomenological equation used to describe dissipative spin systems. Hence, the generalized LL equation~\eqref{MagnetizationDynamics1} for the magnetization vector is capable of reproducing different models. Also, it is closely connected to a Markovian master equation in the mean-field approximation and stands as one of the most relevant results in this work. Moreover, our microscopic model shows that a suitable choice of the common boson bath~\eqref{MeanFiedlState} can generate a particular Lindblad operator $\mathcal{L}_{\rm nn}(\rho)$~\eqref{MasterEquation} which is intimately related to the damping of the precession of the magnetization vector. In the next section, we will explore the magnetic properties of the system by analyzing different hysteresis curves. \par

\begin{figure*}[ht]
\centering
\includegraphics[width=0.8 \linewidth]{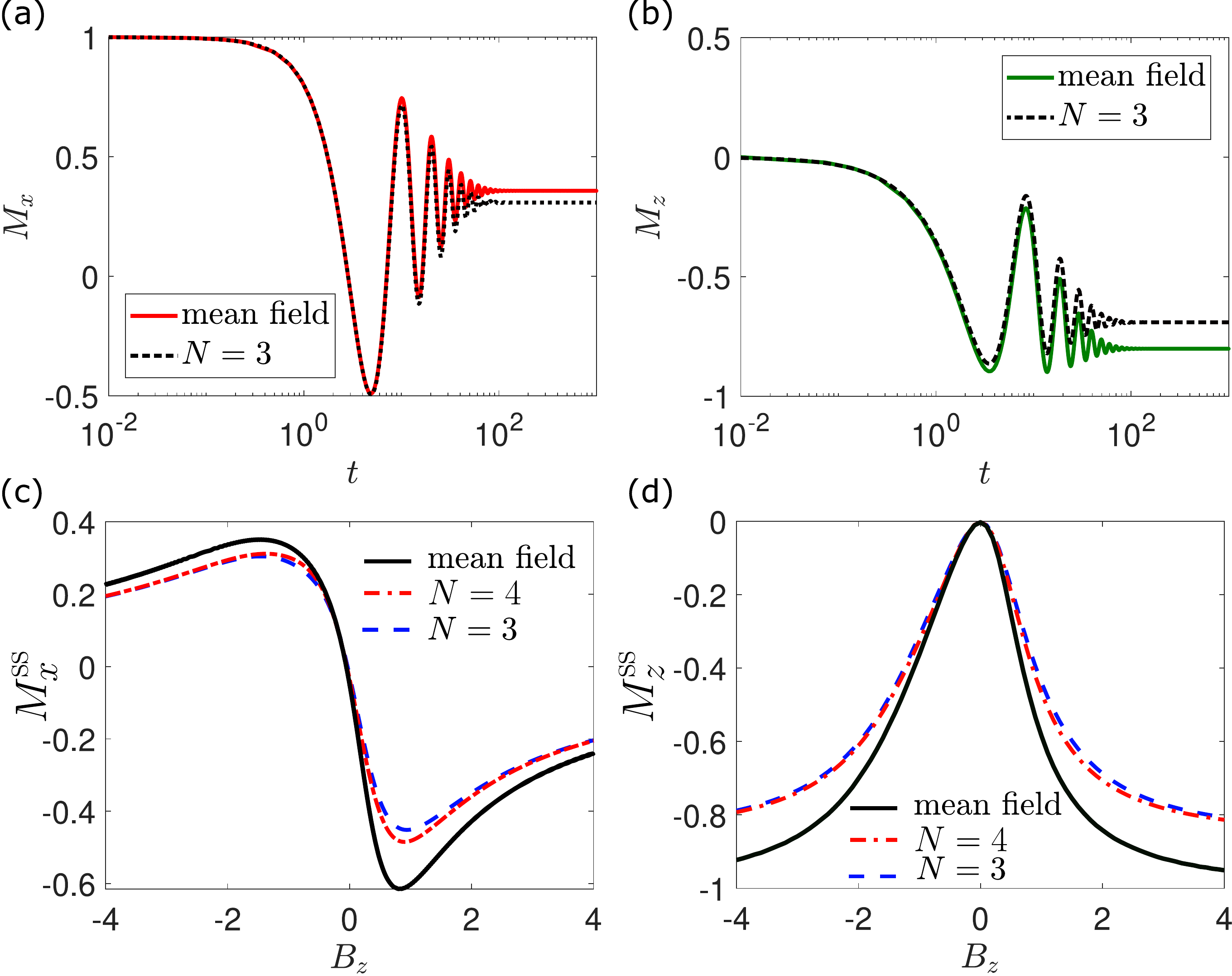}
\caption{Comparison between the mean-field theory and the exact numerical simulation for the magnetization components: (a) $M_x(t)$, (b) $M_z(t)$ using $B_{x,y} = 0.25$, $B_z = -0.5$, $V_x/5 = V_y = V_z =  0.1$, and $\Gamma = 0.1$. (c) and (d) show the steady state for the transverse ($M_x$) and longitudinal ($M_z$) components as a function of the magnetic field component $B_z$ using the mean-field and exact numerical calculations. For the calculations of the steady states we use $B_{x,y}=1$, $V_{x}= V_y/2 = V_z =1$, and $\Gamma = 0.1$.}
\label{fig:Figure3}
\end{figure*} 

\subsection{Hysteresis curves} 

Systems with hysteresis are relevant in nature because they can be experimentally manipulated to understand their response under an external force or action~\cite{Krasnoselskii1989}. In magnetic materials, the hysteresis curve is the relation between the steady state (SS) of the magnetization as a function of the external magnetic field applied along an arbitrary direction. As a first step, we neglect the on-site dissipation ($\Gamma = 0$) and thus, the magnitude of $\mathbf{M}$ is constant during the dynamics. Moreover, we set $\mathbf{D} = \alpha \mathbf{B}_{\rm eff}$ with $\alpha <1$ in order to model the LL dynamics. The SS of the magnetization, $d\mathbf{M}^{\rm SS}/dt = 0$, can be obtained by solving $\mathbf{M}^{\rm SS} \times \mathbf{B}_{\rm eff} + \alpha \mathbf{M}^{\rm SS} \times (\mathbf{M}^{\rm SS} \times \mathbf{B}_{\rm eff}) = 0$, and the non-trivial solution imposes that $\mathbf{M}^{\rm SS}$ must be parallel to the effective magnetic field. Under these assumptions, the dynamics is described by the LL equation~\eqref{LLG}, where the effective magnetic field is $\mathbf{B}_{\rm eff} = \mathbf{B} +\mathbf{B}_{\rm an}$. \par 

We numerically solve the time evolution of the system for the parametrized initial condition $\mathbf{M}(0) = |\mathbf{M}(0)|(\cos(\phi_0)\sin(\theta_0),\sin(\phi_0)\sin(\theta_0),\cos(\theta_0))$. For simplicity, we choose the initial angles as $\phi_0 = 0$ and $\theta_0 = \pi/40$ to simulate a magnetic system slightly misaligned respect to the $z$-axis. In what follows, we introduce our natural units by setting $\gamma = 1$ and $|\mathbf{M}(0)|=1$. As a consequence, the magnetization components satisfy $|M_\alpha(t)| \leq 1$ for $t \geq 0$. In Fig.~\ref{fig:Figure2}(a) we show the time evolution $M_{\alpha}(t)$ for an effective magnetic field $\mathbf{B}_{\rm eff} = (B_z + m_z V_z)\mathbf{e}_z$, with $B_z = -2$, $V_z = 0.5$, and considering $N=500$ spins. One can observe that $\mathbf{M}$ undergoes a dissipative precession leading to $\mathbf{M}^{\rm SS} = (0,0,-1)$. As initially $M_z(0)\approx 1$, the anisotropy field at $t=0$, $\mathbf{B}_{\rm an}(0) = m_z(0) V_z \mathbf{e}_z$, points in the $\mathbf{e}_z$ direction. Then, in the presence of a negatively increasing magnetic field $B_z<0$, the longitudinal component $B_{{\rm eff},z} =B_z+m_zV_z$ becomes negative below the critical magnetic field $B_z^{\rm crit} = -m_z V_z$, meaning that now $\mathbf{B}_{\rm eff}$ points in the $-\mathbf{e}_z$ direction. Thus, as the magnetization follows $\mathbf{B}_{\rm eff}$ in order to reach the SS, all the spins suddenly rotate and the system ends in the final state $M_z(\infty) = -1$. Evidently, for a larger anisotropy field $V_z$ is necessary a larger negative component of the external magnetic field $B_z$ to generate this collective rotational effect. The situation is reversed when $M_z(0)\approx -1$, i.e $\mathbf{B}_{\rm an}(0)$ points in the $-\mathbf{e}_z$ direction, and thus a positive external field $B_z>m_zV_z$ is necessary to induce the rotation towards the stationary state $M_z(\infty)=1$. These observations explain the hysteresis curves illustrated in Fig.~\ref{fig:Figure2}(b). \par

Now, we investigate the effect of including an additional perpendicular field $\mathbf{B}_{\perp} = B_x \mathbf{e}_x+B_y \mathbf{e}_y$. We choose $B_x = B_y = 1$ but preserving $V_x = V_y = 0$. In Fig.~\ref{fig:Figure2}(c) we show the time evolution of the components $M_\alpha(t)$ under the effect of $\mathbf{B} = \mathbf{B}_{\parallel}+\mathbf{B}_{\perp}$, with $\mathbf{B}_{\parallel} = B_z \mathbf{e}_z$. The inclusion of the perpendicular field results in a SS with perpendicular components, i.e. $\mathbf{M}^{\rm SS} = (0.31,0.31,-0.89)$. As a consequence, the hysteresis curves in Fig.~\ref{fig:Figure2}(d) have a smooth dependence in terms of the external field $B_z$, and in some cases ($V=0.5$ and $V=1$), the magnetic coercivity (width of the hysteresis curve) is zero because the anisotropy is weaker than the in-plane applied field $(B_x^2 + B_y^2)^{1/2}=\sqrt{2}$. Note that the steep transitions observed in Fig.~\ref{fig:Figure2}(b) are due to a field that is collinear with the anisotropy. When strong $x,y$ components of the magnetic field are present, the anisotropy becomes less relevant within $\mathbf{B}_{\rm eff}$ and $\mathbf{M}$ follows more readily the direction of $\mathbf{B}$. \par

Our hysteresis curves are in agreement with those studied using the Stoner-Wohlfarth theory~\cite{Atherton1990,Tannous2008,Stoner1948}, which minimizes the energy of a magnetic system $E = -\mathbf{B}\cdot\mathbf{M} - V_z M_z^2$ using either a single domain description or a mean-field approximation. In the next section, we solve the spin dynamics beyond the mean-field approximation by numerically solving the master equation.

\section{Density matrix formalism} \label{Density-Matrix}

In this section, we compare the mean-field model~\eqref{MagnetizationDynamics1} with the numerical solution of the master equation~\eqref{MasterEquation}. Let's begin with a different initial condition for the normalized magnetization, say $\mathbf{M}(0) = (1,0,0)$ or equivalently all spins in the state $\rho_j^{x} = \ket{\uparrow_x}\bra{\uparrow_x}$ with $\ket{\uparrow_x} = (\ket{\uparrow_z}+ \ket{\downarrow_z})/\sqrt{2}$, where $\ket{\uparrow_z}$ and $\ket{\downarrow_z}$ are the eigenstates of $\sigma_z$. The magnetization components $M_{\alpha} = (1/N)\sum_{j}\langle \sigma_\alpha^{(j)} \rangle$ are immediately computed from the density matrix $\rho(t)$ using the relation $\langle \sigma_\alpha^{(j)} \rangle = \mbox{Tr}[\sigma_\alpha^{(j)} \rho]$. A detailed numerical method to solve the master equation is given in the Appendix~\eqref{appendix1}, where the implementation of the Lindblad superoperator $\mathcal{L}_{\rm on}(\rho)$ is presented. In Fig.~\ref{fig:Figure3}(a),(b) we show the time evolution of the magnetization components $M_x(t)$ and $M_z(t)$ for an effective magnetic field $\mathbf{B}_{\rm eff}$ with $B_{x,y} = 0.25$, $B_z = -0.5$, and $V_{x}/5=V_{y}=V_{z} =  0.1$. We observe that the mean-field model partially recover the dynamics of the exact density matrix approach, exhibiting a good agreement at shorter times. \par

The main mismatch occurs at longer times, i.e., the prediction of the stationary state of the system. We remark that the assumption of the mean-field approximation is the product decomposition given in Eq.~\eqref{MeanFiedlState}. However, it is expected that local interactions between spins governed by $V_\alpha$~\eqref{SystemHamiltonian} and the nearest-neighbor Lindbladian $\mathcal{L}_{\rm nn}$~\eqref{MasterEquation} could generate correlations during the dynamics, even starting from an uncorrelated many-body state. The latter is discussed in Sec.~\ref{Correlations}, where correlations are analyzed in more detail. In Fig.~\ref{fig:Figure3}(c),(d) we show the stationary states $M_x^{\rm SS}$ and $M_{z}^{\rm SS}$ as a function of the applied magnetic field $B_z$ for $N=3$ and $N=4$ spins. The mean-field model recover the non-monotonic shape of the components $M_x^{\rm SS}$ and $M_{z}^{\rm SS}$. We observe that differences between the mean-field and master equation increases as the external magnetic field depart from $B_z = 0$. This can also be understood in terms of the propagation of correlation by the dynamics itself. \par 

For instance, if correlations are created at time 
$\tau > 0$ the density matrix immediately takes the form $\rho(\tau) = \rho_1(\tau)\otimes ... \otimes \rho_N(\tau) + \rho_{\rm corr}(\tau)$, where $\rho_{\rm corr}(\tau)$ accounts for such correlations. Consequently, the open dynamics will be strongly affected by the extra Liouvillian generator $\mathcal{L}(\rho_{\rm corr}) = -i[H,\rho_{\rm corr}(\tau)]+\mathcal{L}_{\rm on}(\rho_{\rm corr}(\tau)) + \mathcal{L}_{\rm nn}(\rho_{\rm corr}(\tau))$.  Hence, for a large magnetic field $B_z$ the propagation of correlations will be dominated by the Hamiltonian contribution on the Lindblad superoperator, i.e the term $-i[B_z \sum_{j=1}^{N}S_z^{(j)},\rho_{\rm corr}(\tau)]$. As a consequence, the stationary state will be affected by these additional corrections neglected by the mean-field model, explaining the differences observed in Fig.~\ref{fig:Figure3}(c),(d). In the next subsection, we discuss in more detail the effect of the spin correlations on the dynamics by considering both closed and open dynamics. \par

\subsection{Spin correlations}\label{Correlations}

In order to explain the mismatch between the mean-field approach and master equation, we remark that the mean-field approximation considers that the density matrix of the system can be written as a separable tensor product, as given in Eq.~\eqref{MeanFiedlState}. Therefore, we state that whenever the system departs from this representation, i.e. correlations between the spins show up, the mean-field theory will be deteriorate. To quantify spins correlations, we introduce the two-point correlation function~\cite{Schachenmayer2015},

\begin{equation} \label{CorrelationFunction}
C^{i j}_{\alpha\beta} = \left \langle \sigma_{\alpha}^{(i)}\sigma_{\beta}^{(j)} \right\rangle - \left\langle \sigma_{\alpha}^{(i)} \right\rangle  \left\langle \sigma_{\beta}^{(j)}\right\rangle,
\end{equation}

where $\lbrace i,j\rbrace = 1,...,N$ are the spin indexes, $\sigma_{\alpha}^{(i)}$ is the 
$i$-th Pauli operator. Typically, the expectation values $\langle \sigma_{\alpha}^{(i)} \rangle$ are calculated by assuming the system at thermal equilibrium. However, in our microscopic model we are interested in the non-equilibrium properties of the system. Therefore, we calculate the expectation values as $\langle \sigma_{\alpha}^{(j)} \rangle = \mbox{Tr}[\sigma_{\alpha}^{(j)}\rho_j(t)]$ with $\rho_j(t) = \mbox{Tr}_{1,...,N \neq j}[\rho(t)]$ with $\rho(t)$ being the solution of the master equation~\eqref{MasterEquation}. For further comparison, we denote $\mathbf{M}^{\rm MF}$ as the mean-field solution of the magnetization vector which is obtained by solving Eqs.~\eqref{x}-\eqref{z}. In parallel, we compute $\mathbf{M}^{\rm Exact}$ from the exact master equation~\eqref{MasterEquation} using the definition given in Eq.~\eqref{Magnetization}. Differences between both approaches will quantified in terms of $|\mathbf{M}^{\rm MF}-\mathbf{M}^{\rm Exact}|^2$. In the next subsections we analyse two different scenarios, Case I: $\Gamma = g_{\alpha} = 0$ (closed dynamics) and Case II: $\Gamma,g_{\alpha} > 0$ (Markovian open dynamics)

\subsubsection{Case I: $\Gamma = g_{\alpha} = 0$}\label{CaseI}

For a system described by a closed dynamics, ($\Gamma = g_{\alpha} = 0$), the time evolution of the spin chain is governed by the Heisenberg Hamiltonian~\eqref{SystemHamiltonian}. To better understand the role of correlations, it is instructive to distinguish between the \textit{isotropic} and \textit{anisotropic} Heisenberg models. A fully isotropic Heisenberg model occurs when $V_x = V_y = V_z = V$. Now, we shall illustrate that isotropic interactions are connected with hidden Hamiltonian symmetries. Thus, we define the isotropic interaction Hamiltonian

\begin{equation}
H^{\rm iso} = V\sum_{\alpha = x,y,z}\sum_{j=1}^{N-1}S_{\alpha}^{(j)}S_{\alpha}^{(j+1)},
\end{equation}

which is the second term of Hamiltonian~\eqref{SystemHamiltonian} for $V_{\alpha} = V$. To take into account conserved quantities, we introduce the total spin operator for each component

\begin{equation}
S_{\alpha}^{\rm tot} = \sum_{j =1}^{N}S_{\alpha}^{(j)}.
\end{equation}

We note that the above operator is related to the components of the magnetization vector~\eqref{Magnetization} through the relation $M_{\alpha} = 2\mu \langle S_{\alpha}^{\rm tot}\rangle/(VN \hbar)$. We get $[H^{\rm iso},S_{\alpha}^{\rm tot}]=0$ for each component $\alpha = x,y,z$ which is known as the $SU(2)$ symmetry. Hence, the total spin operator is a conserved quantity under the action of $H^{\rm iso}$, which means that the Hilbert space of the $N$ particles separates into disjunct Hilbert subspaces with constant magnetization. As the Zeeman contribution on the Heisenberg Hamiltonian~\eqref{SystemHamiltonian} is a local effect, we conclude that spin correlations cannot be generated for the isotropic case if the system does not interact with an external environment. These observations implies that the mean-field decomposition~\eqref{MeanFiedlState} is valid for the isotropic case if $\Gamma = g_{\alpha} = 0$. As a consequence, the two-point correlation functions $C^{i j}_{\alpha\beta} = 0$ for all $ i,j,\alpha,\beta$, revealing that the mean-field and exact models are identical in this particular case (which is also numerically corroborated). In other words, in the absence of a reservoir, the only non-local effect comes from the coupling $V_{\alpha}S_{\alpha}^{(j)}S_{\alpha}^{(j+1)}$, and spin correlations only arise when we depart from the ideal isotropic case $V_{\alpha}=V$. \par

\begin{figure}
\centering
\includegraphics[width=0.8 \linewidth]{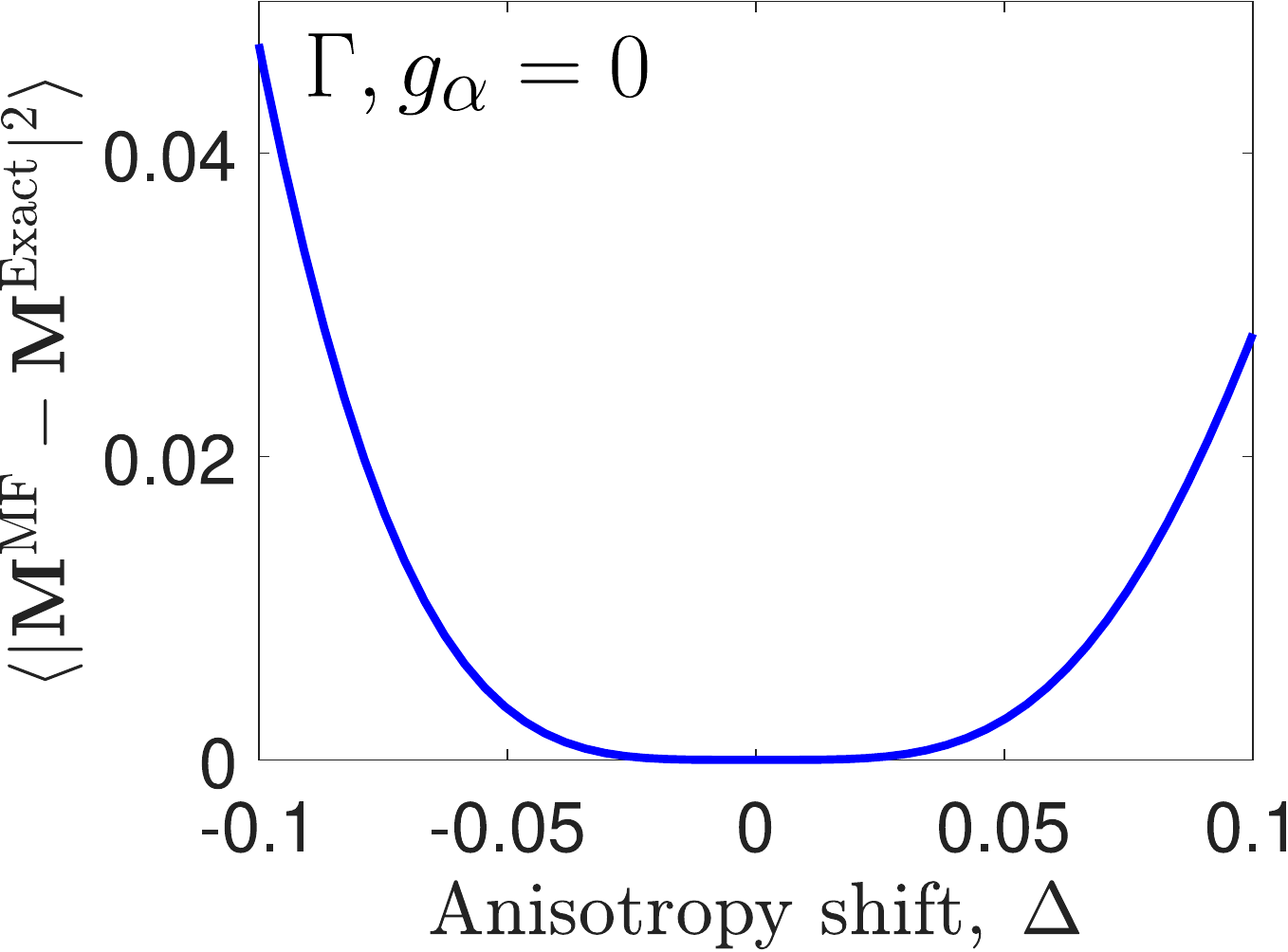}
\caption{Time average of $|\mathbf{M}^{\rm MF}-\mathbf{M}^{\rm Exact}|^2$ as a function of the anisotropy shift $\Delta$. For the simulation we consider three spins with $V_x = V+\Delta$, $V_y = V_z = V$ with $V = 0.1$ and an external magnetic field $B_x = B_y = 0.25$ and $B_z = -0.5$. For the initial condition we consider three spins aligned in the $x$ direction. Here, $\mathbf{M}^{\rm MF}$ and $\mathbf{M}^{\rm Exact}$ are the mean-field and exact magnetization vectors, respectively.}
\label{fig:Figure4}
\end{figure} 

Let us consider a spin chain with anisotropy such that $V_x = V + \Delta$ and $V_y = V_z = V$, where $\Delta$ and $V$ are the anisotropy shift and the isotropic contribution, respectively. For $\Delta = 0$, we recover the previous case where non-local interactions leaves invariant the magnetization vector. For $\Delta >0$ the anisotropy is present along the $x$ axis, while for $\Delta <0$ the anisotropy is changed to the $y, z$ plane. In addition, we use the same initial condition detailed in Sec.~\ref{Density-Matrix}, where $\mathbf{M}(0) = (1,0,0)$ and we fix $V = 0.1$. In Fig.~\ref{fig:Figure4} we plot the time average of $|\mathbf{M}^{\rm MF}-\mathbf{M}^{\rm Exact}|^2$ for a time interval $(0,10^3)$ by considering $B_x = B_y = 0.25$, $B_z = -0.5$, and $N=3$. We observe that $\langle |\mathbf{M}^{\rm MF}-\mathbf{M}^{\rm Exact}|^2\rangle$ monotonically increases with the anisotropy shift. This means that configurations without $SU(2)$ symmetry ($\Delta \neq 0$) can generate correlations between spins and $\mathbf{M}^{\rm SS}\neq\mathbf{M}^{\rm Exact}$ in such cases. In the next subsection, we analyse the effect of including the losses induced by the bosonic thermal environment.

\subsubsection{Case II: $\Gamma, g_{\alpha} > 0$}

\begin{figure*}[ht]
\centering
\includegraphics[width=0.8 \linewidth]{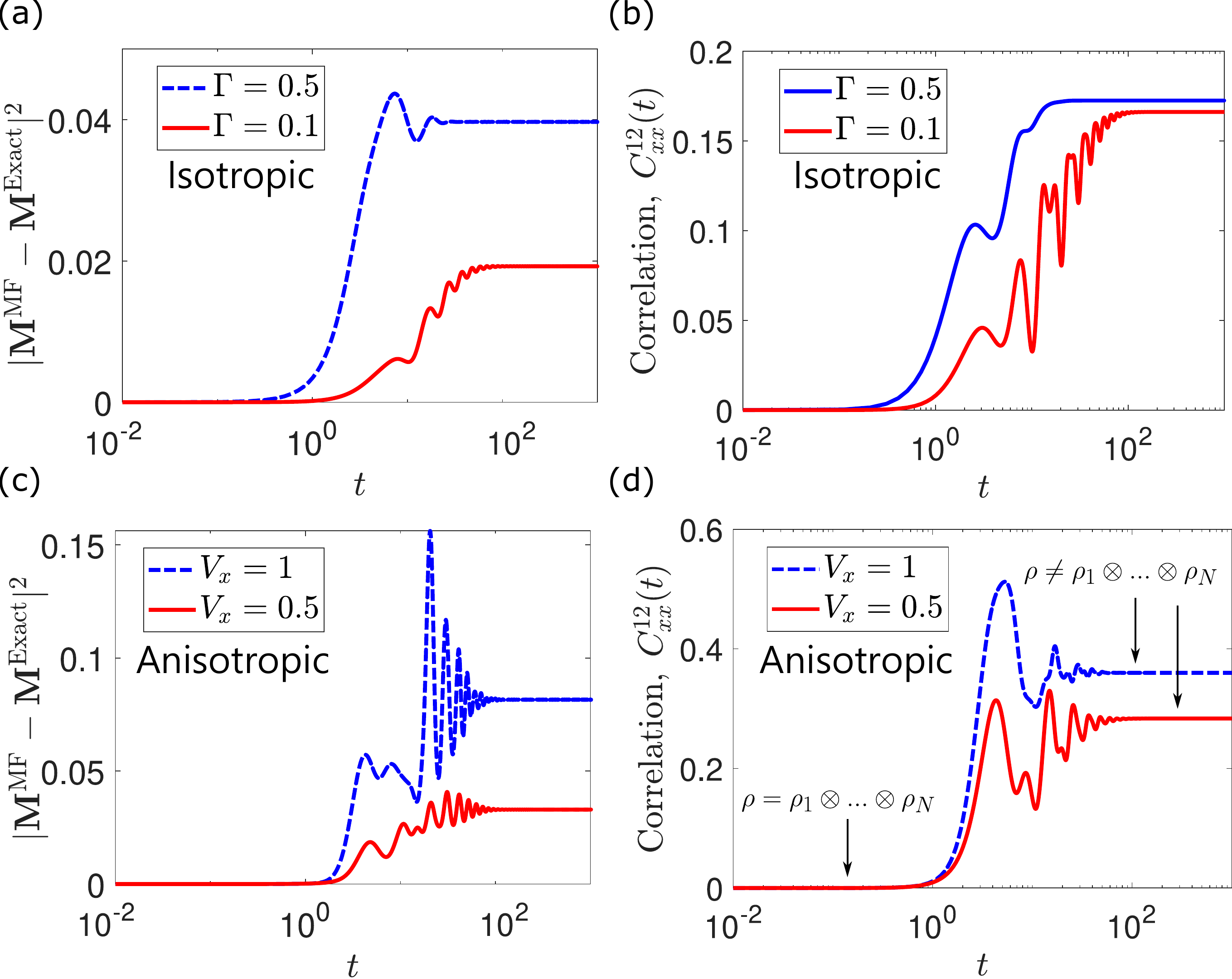}
\caption{Comparison between the mean-field and quantum master equation approaches for the: (a) isotropic and (c) anisotropic open Heisenberg models, respectively. Two-point correlation function $C_{xx}^{12}$ for: (b) isotropic and (d) anisotropic models. General parameters for (a)-(d): $\rho(0) = \rho_1(0) \otimes \rho_2(0) \otimes \rho_3(0)$, where $\rho(0) = |\uparrow_x\rangle\langle \uparrow_x|$ with $|\uparrow_x\rangle = (|\uparrow_z\rangle +|\downarrow_z\rangle)/\sqrt{2}$, $B_{x,y} = 0.25$, $B_z = -0.5$, $V_{y,z} = 0.1$, and  $g_{\alpha} = \Gamma/10$. For (a) and (b) we fix $V_x = 0.5$, and conversely, for (c) and (d) we fix $\Gamma = 0.1$.}
\label{fig:Figure5}
\end{figure*}

For the Markovian open dynamics, ($\Gamma, g_{\alpha} > 0$), the evolution of the system is ruled by the Lindblad master equation~\eqref{MasterEquation}. For an isotropic Heisenberg Hamiltonian, the only source of non-locality is given by the nearest-neighbours Lindbladian $\mathcal{L}_{\rm nn}$, which is by definition a non-local superoperator acting on adjacent spins. In Fig.~\ref{fig:Figure5}(a) we plot the function $|\mathbf{M}^{\rm MF}-\mathbf{M}^{\rm Exact}|^2$ for the isotropic case by considering that all spin are aligned in the $x$ direction at $t=0$. We define $\gamma_{z,+} = \gamma_0 n_b$ and $\gamma_{z,-} = \gamma_0 [n_b + 1]$ with 
$\gamma_0 = \Gamma/(2n_b+1)$ and $n_b$ the mean number of phonons such that $\Gamma = \gamma_{z,-}+\gamma_{z,+}$ for $\mathcal{L}_{\rm on}$ and $g_{z,\eta} = \gamma_{z,\eta}/10$ for $\mathcal{L}_{\rm nn}$. Along this work we set $n_b = 0.08$ for our simulations.

At the beginning of the dynamics, the mean-field and master equation predict the same magnetization vector, see Fig.~\ref{fig:Figure5}(a). However, after a critical time (which is shorter for larger values of $\Gamma$), we observe a deviation between both magnetization vectors, leading to a constant stationary difference at longer times. For comparison, in Fig.~\ref{fig:Figure5}(b) we show the time evolution of the two-point correlation function $C_{xx}^{12}(t)$. For simplicity we only show $C_{xx}^{12}$, however, the same behaviour is numerically obtained for other components, i.e $C_{\alpha \beta}^{12}$ with $\alpha, \beta = x,y,z$. At shorter times we observe that $C_{xx}^{12}(t) = 0$ which means that the density matrix is given by the product state~\eqref{MeanFiedlState}. This result supports the good agreement between the mean-field model and the master equation since the mean-field assumption is fully satisfied. As time increases, the system can no longer be described as a product state leading to $C_{xx}^{12}(t) \neq 0$, and the mean-field approximation fails. We remark that in the limit $\Gamma \rightarrow 0$ the system converges to the isotropic closed dynamics, i.e $\mathbf{M}^{\rm MF} = \mathbf{M}^{\rm Exact}$ for all times. \par

For an anisotropic Heisenberg Hamiltonian, we have two sources of non-locality, the interaction Hamiltonian (dominant) and the Lindbladian $\mathcal{L}_{\rm nn}$ (small contribution). Now, we simulate the previous case but adding anisotropy in the $x$ direction, i.e. $V_y = V_z = 0.1$ and $V_x =  \{0.5,1\}$. In Fig.~\ref{fig:Figure5}(c), we observe that the addition of the anisotropy increases the mismatch between the mean-field and exact calculations. Furthermore, as one would expect, the coupling constant $V_x$ (local interactions between spins) contributes to increasing correlations, which is illustrated in Fig.~\ref{fig:Figure5}(d). Moreover, we observe that the mean-field model can not exactly predict the steady-state of the system because of the correlations originated by non-local terms, which confirms our previous observations in Fig.~\ref{fig:Figure3}.  \par

\subsubsection{Quantum correlations}

To complement the previous analysis based on correlations we note that one remaining question is whether the observed correlations are quantum in nature or just a statistical mixing of the density matrix, i.e. a mixed state. Quantum correlations (QC) are one of the most fundamental concepts in Quantum Information Theory~\cite{FanchiniBook}. Even more, QC provides a useful resource to speed up several tasks in quantum computing~\cite{Alastair2015}. In particular, the \textit{concurrence} defined by Wootters~\cite{Wootters} in the context of a two-qubit system, is a widely used measure that account for QC based on the separability of the system. For a system with $N$ spins we trace over $N-2$ spins and we obtain the two-qubit density matrix $\rho_{ij}(t)=\mbox{Tr}_{1,...,N \neq i,j}[\rho(t)]$, where particles $i$ and $j$ are two different arbitrary particles of the spin chain. Then, we calculate the \textit{spin-flipped} density matrix $\tilde{\rho}_{ij}=(\sigma_y^{(i)}\otimes\sigma_y^{(j)})\rho_{ij}^{\ast}(\sigma_y^{(i)}\otimes\sigma_y^{(j)})$, and the Concurrence is given by $\mathcal{C}_{ij}=\max{\lbrace0,\alpha_1-\alpha_2-\alpha_3-\alpha_4\rbrace}$, where the $\alpha_1,\,\alpha_2,\,\alpha_3,\,\alpha_4$ are the square root of the eigenvalues of $\rho_{ij}\tilde{\rho}_{ij}$ in decreasing order. \par

\begin{figure}[ht]
\centering
\includegraphics[width=0.8 \linewidth]{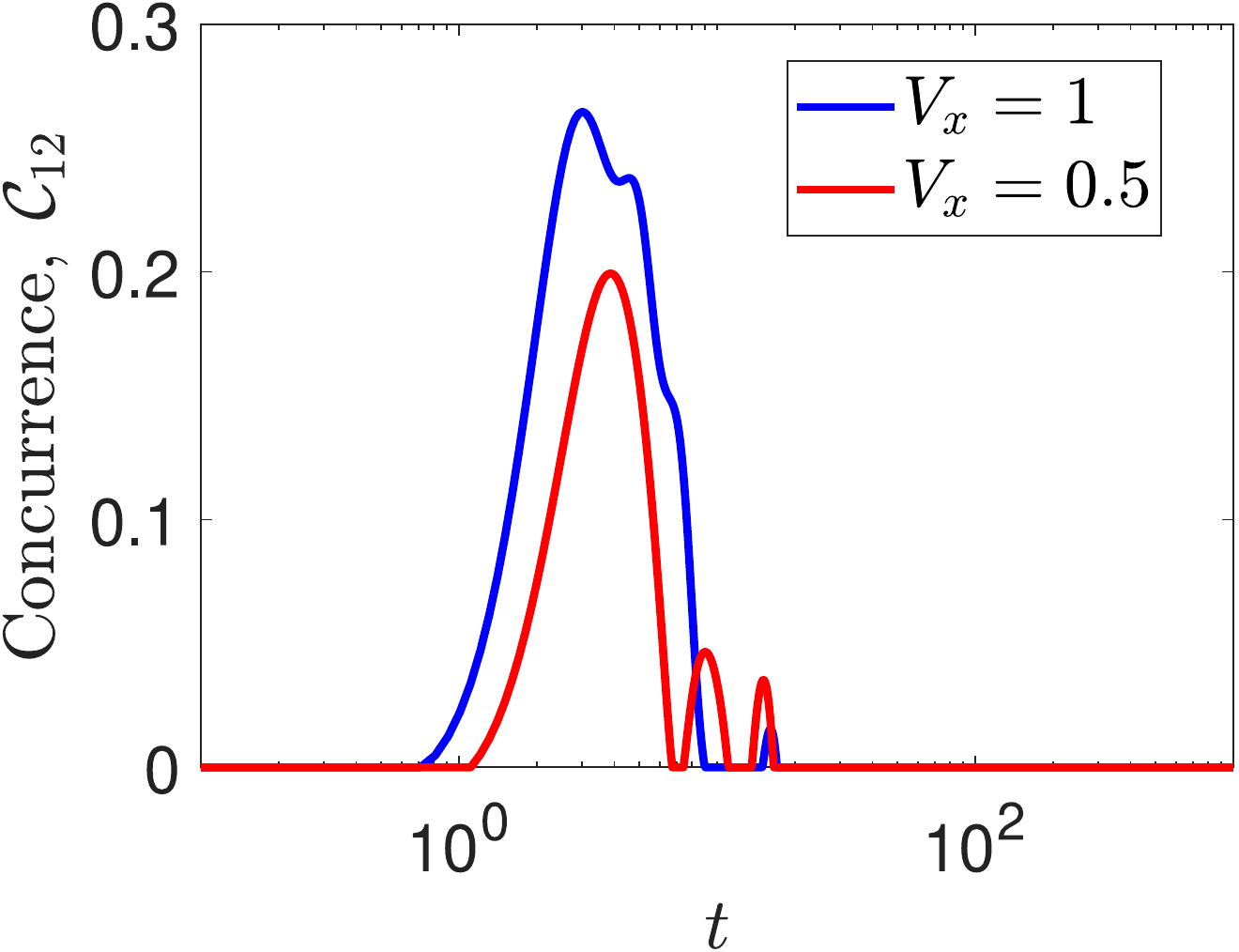}
\caption{Concurrence $\mathcal{C}_{12}$ as a function of three spins in linear configuration. For the simulation we consider all spins aligned in the $x$ direction at $t = 0$, with $V_y = V_z = 0.1$, $B_{x,y} = 0.25$, $B_z = -0.5$, $\Gamma = 0.1$, and $g_{\alpha} = \Gamma/10$. }
\label{fig:Figure6}
\end{figure}

The concurrence is a bounded function, $0 \leq \mathcal{C}_{ij} \leq 1$, where $\mathcal{C}_{ij} = 0$ means zero entanglement between particles $i$ and $j$. In Fig.~\ref{fig:Figure6} we plot the concurrence $\mathcal{C}_{12}$ for $N=3$ by considering the open anisotropy case illustrated in Fig.~\ref{fig:Figure5}(c),(d). First, we observe that QC reaches a non-negligible maximum value, i.e $\mbox{max}[\mathcal{C}_{12}] \approx 0.26$ for $V_x = 1$. However, for the parameters used in the simulation, we note that QC appears in a short time window, which is related to the temporal region where both master equation and mean-field models are different. Second, QC increases with the local coupling term $V_x$, which is expected since the interaction between adjacent spins creates bi-partite entangled states. More generally, one could find QC by performing a local measurement on the spins instead of tracing out them. This allows us to define a localizable entanglement~\cite{Verstraete}, which has been shown that is closely related to the two-point correlation function~\cite{Verstraete}. Nevertheless, local measurements will involve additional resources that we are not considering here, and thus, it is out of our scope. \par

\section{Conclusions} 

In summary, we established an unexplored microscopic connection between the open Heisenberg model and the Landau-Lifshitz equation. Starting from a generic spin-boson interaction Hamiltonian, we derived a Markovian master equation, and applying the mean-field approximation, we found a generalized LL equation. Consequently, we recognized the microscopic origin of anisotropy effects, dissipative magnetic fields, and relaxation processes induced by the Heisenberg Hamiltonian and external boson bath. First, we focused on the hysteresis curves for longitudinal and transverse magnetic fields, reaching a good agreement with the Stoner-Wohlfarth theory. We also solved the non-equilibrium dynamics by numerical calculations of the master equation for a small number of spins. We compared the mean-field and master equation, revealing that a mean-field model phenomenologically describes the main magnetic features such as temporal behavior (oscillations and decay) and stationary states as a function of external fields, although with some deviations on their exact behavior. \par

Using the two-point correlation function and the concurrence, we showed that these deviations are due to a correlation originated from magnetic anisotropy and the non-local Lindbladian between spins, making the central assumption of the mean-field approximation invalid. We expect these deviations to be negligible as the number of spins increases. Finally, our model can be used to connect the dynamics of open quantum systems with magnetic-like systems.

\section{acknowledgments} AN acknowledges financial support from Universidad Mayor through the Postdoctoral fellowship. RC acknowledges financial support from Fondecyt Iniciaci\'on No. 11180143. 

\appendix

\section{Derivation of the master equation}\label{appendix2}

In the spin basis $\ket{\uparrow}_z^{(j)}$ and $\ket{\downarrow}_z^{(j)}$ (eigenstates of $\sigma_z^{(j)}$), the states $\ket{\downarrow}_{\alpha}^{(j)}$ are given by

\begin{eqnarray}
\ket{\uparrow_x}^{(j)} &=& {1 \over \sqrt{2}}\left(\ket{\uparrow}_z^{(j)}+\ket{\downarrow}_z^{(j)}\right), \label{xup} \\
\ket{\downarrow_x}^{(j)} &=& {1 \over \sqrt{2}}\left(-\ket{\uparrow}_z^{(j)}+\ket{\downarrow}_z^{(j)}\right), \\
\ket{\uparrow_y}^{(j)} &=& {1 \over \sqrt{2}} \left(\ket{\uparrow}_z^{(j)}+\mbox{i}\ket{\downarrow}_z^{(j)} \right), \\
\ket{\downarrow_y}^{(j)} &=& {1 \over \sqrt{2}} \left(-\ket{\uparrow}_z^{(j)}+\mbox{i}\ket{\downarrow}_z^{(j)}\right). \label{zdown}
\end{eqnarray}




To derive the open dynamics of the spin chain, we move to the interaction picture, where the Liouville-Von Neumann equation read as ($\hbar = 1$)



\begin{eqnarray}\label{MasterEquation1}
{d \tilde{\rho}_s \over dt} &=& -i\mbox{Tr}_{b}\left([\tilde{V}(t),\tilde{\rho}_{s+b}(0)]\right)\nonumber  \\
&& -\int_{0}^{t} dt' \mbox{Tr}_b\left([\tilde{V}(t),[\tilde{V}(t'),\tilde{\rho}_{s+b}(t')]]\right),
\end{eqnarray}

where $\tilde{\rho}_s = \mbox{Tr}_{b}[\tilde{\rho}_{s+b}(t)]$ is the reduced density matrix in the interaction picture, $\tilde{\rho}_{s+b}(t) = \mbox{exp}(-i H_0 t)\rho_{s+b}(t)\mbox{exp}(i H_0 t)$ with $H_0 = H_s + H_b$ accounting for both Heisenberg and phonon Hamiltonians. At thermal equilibrium, the bath density matrix is $\rho_b = \mbox{exp}(-\beta H_b)/Z$, where $Z = \mbox{Tr}_{b}(\mbox{exp}(-\beta H_b))$ is the partition function, $\beta = (k_B T)^{-1}$ is the inverse temperature, and $H_b = \sum_k \omega_{k}a_k^{\dagger}a_k$ is the bath Hamiltonian (harmonic oscillators). In addition, we employ the born approximation~\cite{Vega2017}, where is assumed that at any time the full density matrix can be decomposed as an uncorrelated product state, i.e. $\tilde{\rho}_{s+b}(t) = \tilde{\rho}_s(t) \otimes \rho_b$. The latter is valid in the \textit{weak-coupling limit}, which is fulfilled when $g_{\alpha jk} \ll \mbox{max}(\gamma |\mathbf{B}|,V_{\alpha})$. Under these assumptions, we have $\mbox{Tr}_b(a_k \rho_b) = \mbox{Tr}_b(a_k^{\dagger} \rho_b) = 0$, and therefore the first term of the right-hand of Eq.~\eqref{MasterEquation1} vanishes. As a consequence, we derive the following convolution dynamics

\begin{eqnarray}\label{MasterEquation2}
{d \tilde{\rho}_s \over dt} &=& \sum_{\alpha, j;\alpha' j'} \sum_k \int_{0}^{t} d\tau  A_{\alpha j;\alpha' j'}^{k} e^{i\omega_k \tau} \times \nonumber \\
&& \left[S_{\alpha',+}^{(j')}(t-\tau) \tilde{\rho}_s(t') S_{\alpha,-}^{(j)}(t)
 - S_{\alpha,-}^{(j)}(t) S_{\alpha',+}^{(j')}(t-\tau)\tilde{\rho}_s(t')  \right] \nonumber \\
&+& \sum_{\alpha,j;\alpha'j'} \sum_k \int_{0}^{t} d\tau  B_{\alpha j;\alpha' j'}^{k} e^{-i\omega_k \tau} \times \nonumber \\
&& \left[S_{\alpha',-}^{(j')}(t-\tau) \tilde{\rho}_s(t') S_{\alpha,+}^{(j)}(t) - S_{\alpha,+}^{(j)}(t) S_{\alpha',-}^{(j')}(t-\tau)\tilde{\rho}_s(t') \right] \nonumber \\
&& +h.c,
\end{eqnarray}

where $A_{\alpha j;\alpha'j'}^{k} = g_{\alpha jk}^{\ast} g_{\alpha' j'k}n(\omega_k)$, $B_{\alpha j;\alpha' j'}^{k} = g_{\alpha jk} g_{\alpha' j'k}^{\ast}[n(\omega_k)+1]$ are the coupling terms associated with absorption and emission processes, respectively. Here, $n(\omega_k) = [\mbox{exp}(\hbar \omega_k/k_B T)-1]^{-1}$ is the mean number of bosons at thermal equilibrium. Now, we applied the \textit{first and second Markov approximations}~\cite{Vega2017} and we assume that $\tilde{\rho}_s(t') \approx \tilde{\rho}_s(t)$ and that the integral contribution can be evaluated at larger times, i.e. for $t \rightarrow \infty$. Now, we introduce the spectral decomposition~\cite{Breuerbook,Stanislaw2008} 

\begin{eqnarray}
S_{\alpha,\pm}^{(j)} &=& \sum_{\omega}S_{\alpha,\pm}^{(j)}(\omega), \\
S_{\alpha,\pm}^{(j)}(\omega) &=& \sum_{a,b}\delta(\omega_{ba}-\omega) |a\rangle \langle a|S_{\alpha,\pm}^{(j)} |b\rangle \langle b|,
\end{eqnarray}

where $\delta(\omega_{ba}-\omega)$ is a Kronecker function, i.e. $\delta(x)= 1 $ for $x=0$, and $\delta(x)=0$ otherwise. The quantum states $\ket{a},\ket{b}$ are eigenstates of the Heisenberg Hamiltonian~\eqref{SystemHamiltonian}, with $V_{\alpha \beta}^{ij} = \delta_{\alpha \beta} \delta_{i,j-1}V_{\alpha}$ and $\alpha = \beta$. In the interaction picture, the following relations are satisfied in the frequency domain 

\begin{eqnarray} 
S_{\alpha,-}^{(j)}(t) &=& \sum_{\omega}e^{-i \omega  t} S_{\alpha,-}^{(j)}(\omega), \label{SpectralDecomposition1} \\
S_{\alpha',+}^{(j')}(t') &=& \sum_{\omega'}e^{i \omega'  t'} S_{\alpha',+}^{(j')}(\omega'). \label{SpectralDecomposition2}
\end{eqnarray}

By replacing the operators $S_{\alpha,-}^{(j)}(t)$ and $S_{\alpha',+}^{(j')}(t')$ into Eq.~\eqref{MasterEquation2} using the spectral decomposition~\eqref{SpectralDecomposition1} and~\eqref{SpectralDecomposition2} we obtain the oscillating functions $\mbox{exp}(\pm i (\omega'-\omega)t)$. In the \textit{secular approximation}, we neglect the terms $\omega \neq \omega'$ due to the condition $\tau_b \gg T_s$, where $\tau_b \sim g_{\alpha jk}^{-1}$ and $T_s \sim  1/\max(\gamma |\mathbf{B}|,V_{\alpha}) \sim T_s$ are the bath and system characteristic times, respectively. Therefore, in the secular and Markov approximations, we obtain the following Lindblad master equation in the 
Schr\"{o}dinger picture


\begin{eqnarray}\label{MasterEquation3}
{d \rho_s \over dt} &=& -i[H_s,\rho_s] \nonumber \\
&& +\sum_{\omega}\sum_{\alpha,j;\alpha'j'} \gamma_{\alpha j:\alpha' j'}^{+}(\omega) \left[S_{\alpha',+}^{(j')}(\omega) \rho_s S_{\alpha,+}^{(j),\dagger}(\omega) \right. \nonumber \\
&& \left.  -{1 \over 2}\left\{S_{\alpha,+}^{(j),\dagger}(\omega)S_{\alpha',+}^{(j')}(\omega), \rho_s \right \} \right] \nonumber \\
&& +\sum_{\omega}\sum_{\alpha,j;\alpha' j'} \gamma_{\alpha j:\alpha' j'}^{-}(\omega) \left[S_{\alpha',+}^{(j'),\dagger}(\omega) \rho_s S_{\alpha,+}^{(j)}(\omega) \right. \nonumber \\
&& \left.  -{1 \over 2}\left\{S_{\alpha,+}^{(j)}(\omega)S_{\alpha',+}^{(j'),\dagger}(\omega), \rho_s \right \} \right]  
\end{eqnarray}

where the Lamb-shift Hamiltonian has been neglected. The time-dependent rates are defined as

\begin{flalign}
\gamma_{\alpha j:\alpha' j'}^{+}(\omega) &= 2 \mbox{Re}\left[\sum_{k}\int_{0}^{\infty} d\tau \; A_{\alpha j;\alpha' j'}^k e^{i(\omega-\omega_k)\tau}\right], \\
\gamma_{\alpha j:\alpha' j'}^{-}(\omega) &= 2 \mbox{Re}\left[\sum_{k}\int_{\infty}^{t} d\tau \; B_{\alpha j;\alpha' j'}^k e^{-i(\omega-\omega_k)\tau}\right]. 
\end{flalign}

Finally, we make the last approximations to derive the phenomenological master equation presented in Eq.~\eqref{MasterEquation}. First, we consider a nearest-neighbor interaction model to transfer energy between adjacent spins, which implies that we only consider contributions satisfying the condition $|j-j'|=1$. Second, we assume that the anisotropy induced by the common reservoir has the same form as the Heisenberg model presented in Sec.~\ref{Model}, then $\alpha = \alpha'$. Finally, following the Einstein model's spirit for the heat capacity in solid-state physics, we introduce a phenomenological average resonant frequency $\omega \sim \omega_0$ for all spins and therefore $\omega = \omega_0$. Under these assumptions the Lindblad master equation~\eqref{MasterEquation3} reduces to Eq.~\eqref{MasterEquation}.

\section{Solving the master equation}\label{appendix1}

To numerically solve the master equation we adopt the following general solution~\cite{Dominic2016,Katarzyna2016}

\begin{equation} \label{GeneralSolutionMarkovianMasterEquation}
\rho(t) = \sum_{k=1}^{N} c_k e^{\lambda_k t} R_k,
\end{equation}

where $R_k$ and $L_k$ are the right and left eigenmatrices given by the equations $\mathcal{L}(R_k) = \lambda_k R_k$ and $\mathcal{L}^{\dagger}(L_k) = \lambda_k L_k$, respectively. The Lindblad generator $\mathcal{L}$ is defined from the structure of the Markovian master equation $\dot{\rho} = \mathcal{L}(\rho)$. The matrices $R_k$ and $L_k$ must to satisfy the orthonormality condition $\mbox{Tr}(R_k L_{k'}) = \delta_{kk'}$, $c_k = \mbox{Tr}(\rho(0)L_k)$ are coefficients with $\rho(0)$ being the initial state, and $\lambda_k$ the corresponding eigenvalues of the right eigenmatrices $R_k$. For numerical purposes is convenient to sort the eigenvalues $\lambda_k = \lambda_k^{\rm R}+i\lambda_k^{\rm I}$ by choosing $0=\lambda_1^{\rm R} \leq \lambda_2^{\rm R} ... \leq \lambda_2^{\rm N_{\rm d}}$, where $N_{\rm d} = 2^{2N}$ is the number of eigenvalues of the system. The zero eigenvalue $\lambda_1=0$ is related to the stationary state since others eigenvalues with $k>1$ satisfy $\lambda_k^{\rm R}<0$~\cite{RivasHuelgaBook} leading to dissipative terms $\propto e^{-|\lambda_k^{\rm R}|t} \xrightarrow{t\rightarrow \infty} 0$. To compute the matrices $R_k$ and $L_k$ we employ the formalism presented in Ref.~\cite{Nobuyuki2019}, where the strategy is to rewrite the effect of the Lindblad generator $\mathcal{L}$ on a more involved vector space. To this end, the many-body density matrix is mapped to a new vector space as follow

\begin{equation} \label{MAP}
\rho(t) = \sum_{kl} \rho_{kl}|k\rangle \langle l | \mapsto |\rho\rangle\rangle = {1 \over C}\sum_{kl} \rho_{kl} |k,l \rangle \rangle,
\end{equation}

where $|k,l \rangle \rangle = \ket{k} \otimes \ket{l}$ is the new vector basis constructed by the initial vector basis $\ket{k}$ and $C = (\sum_{k,l}|\rho_{kl}|^2)^{1/2}$ is a normalization factor. In this new vector space spawned by the basis $|k,l \rangle \rangle$ the master equation can be rewritten as~\cite{Nobuyuki2019} 

\begin{eqnarray}\label{SolutionVector}
{d |\rho \rangle \rangle \over dt} &=& \mathcal{\hat{L}} |\rho \rangle \rangle \nonumber \\
                          &=& \left[-i\left(H \otimes \mathds{1}- \mathds{1} \otimes H^{T} \right) + \sum_{\eta = \pm} \gamma_{\eta} \mathcal{D}(S_{\pm}) \right]|\rho \rangle \rangle. \nonumber\\
\end{eqnarray}

The operator $\mathcal{\hat{L}}$ is a complex matrix with $2^{2N}\times 2^{2N}$ elements, and

\begin{equation}
\mathcal{D}(S_{\pm}) = S_{\eta} \otimes S_{\eta}^{\ast}
              -{1 \over 2}\left[(S_{\eta}^{\dagger}S_{\eta})\otimes  \mathds{1}
                                        - \mathds{1} \otimes(S_{\eta}^T S_{\eta}^{\ast}) \right],
\end{equation}

is the dissipative term of the Markovian master equation (boson reservoir). The algorithm to solve the open dynamics is quite simple. First one compute the eigenvalues and eigenvectors of $\mathcal{\hat{L}}$ and $\mathcal{\hat{L}}^{\dagger}$ (in the new basis), then using the map~\eqref{MAP} we rewrite the right and left eigenmatrices in the initial Hilbert space, and finally, we employ the general solution given in Eq.~\eqref{GeneralSolutionMarkovianMasterEquation}. \par



\bibliographystyle{unsrt}

\end{document}